\documentclass{elsart}
\usepackage{amssymb}
\usepackage{bbm}   
\usepackage{latexsym} 
\usepackage{epsfig}
\usepackage{color }

\usepackage{mathrsfs} 

\newtheorem{theorem}{Theorem}[section]

\newtheorem{definition}{Definition}[section]

\newcommand{\Proofstart}{\textit{Proof: }}
\newcommand{\Proofend}{ $\square$ }
\def\bd{\begin{definition}}
\def\ed{\end{definition}}
\def\be{\begin{equation}}
\def\ee{\end{equation}}
\def\bea{\begin{eqnarray}}
\def\eea{\end{eqnarray}}
\def\bitem{\begin{itemize}}
\def\eitem{\end{itemize}}
\def\ben{\begin{enumerate}}
\def\een{\end{enumerate}}
\def\bdescribe{\begin{description}}
\def\edescribe{\end{description}}

\def\a{\alpha}
\def\b{\beta}

\def\c{\gamma}
\def\d{\delta}
\def\e{\varepsilon}
\def\l{\lambda}

\def\t{\tau}

\def\th{\theta}

\def\Om{\Omega}

\def\half{\frac{1}{2}}
\def\third{\frac{1}{3}}

\def\ap{{\a'}}
\def\bp{{\b'}}
\def\mup{{\mu'}}

\renewcommand{\cal}{\mathcal}
\def\TD{{\cal{D}}} 

\newcommand{\TCurv}{{\cal{R}}}
\newcommand{\CY}{Y}
\newcommand{\TE}{{\cal{E}}}

\newcommand{\TB}{{\cal{B}}}

\newcommand{\cg}{\mathbf{g}}

\newcommand{\gtilde}{\tilde{g}}

\newcommand{\D}{\nabla}
\newcommand{\tD}{\tilde{D}}

\newcommand{\Dhat}{\hat{\nabla}}
\newcommand{\Dtilde}{\tilde{\nabla}}

\newcommand{\Atilde}{\tilde{A}} 
\newcommand{\Phat}{\hat{P}}

\newcommand{\Ptilde}{\tilde{P}} 

\newcommand{\utilde}{\tilde{u}}
\newcommand{\stilde}{\tilde{\sigma}} 
\newcommand{\ptilde}{\tilde{p}}
\newcommand{\btilde}{\tilde{b}}
\newcommand{\pitilde}{\tilde{\pi}}
\newcommand{\qtilde}{\tilde{q}}
\newcommand{\mutilde}{\tilde{\mu}}
\newcommand{\Ltilde}{\tilde{L}}

\def\rest{\arrowvert} 
\def\implies{\Rightarrow}

\def\mod{\arrowvert}
\def\norm{\Vert}
\def\after{\circ}
\def\sect{\in\Gamma}
\def\Lie{{\cal{L}}}
\def\real{\mathbbm{R}} 

\newcommand{\tensor}[3]{_{#1 \phantom{#2}#3}^{\phantom{#1}#2}}

\def\non{\nonumber}

\begin{document}

\begin{frontmatter}

\title{A global conformal extension theorem \\[1ex] for perfect fluid Bianchi space-times}

\author{Christian L\"ubbe}
\address{Mathematical Institute, University of Oxford, OX1 3LB, Oxford, UK}
\address{Erwin Schr\"odinger Institute, Boltzmanngasse 9, A-1090 Vienna, Austria}
\ead{christian.luebbe@gmail.com}

\author{Paul Tod}
\address{Mathematical Institute, University of Oxford, OX1 3LB, Oxford, UK}
\address{St John's College, OX1 3JP, Oxford, UK}
\ead{tod@maths.ox.ac.uk}

\begin{abstract}
A global extension theorem is established for isotropic singularities in polytropic perfect fluid Bianchi space-times. When an extension is possible, the limiting behaviour of the physical space-time near the singularity is analysed.
\end{abstract}

\begin{keyword}

 Bianchi space-times \sep singularities \sep conformal geometry
\PACS 02.40.-k \sep 04.20.Dw \sep 04.20.Gz \sep 98.80.Jk
\end{keyword}
\end{frontmatter}

\section{Introduction}

In an earlier article \cite{article1} we proved a local conformal extension theorem, establishing curvature conditions along an incomplete conformal geodesic sufficient to permit an extension of the conformal structure. The proof was phrased in the tractor formalism. As was noted at the end of \cite{article1}, one would like to generalise this result to a global extension theorem, extending simultaneously through all singular points. In this article we analyse perfect fluid Bianchi space-times for isotropic singularities. We find that the fluid flow-lines are conformal geodesics, which allows us to apply the local extension theorem of \cite{article1} to any one of them. Then the spatial homogeneity makes it possible to obtain a global extension result.

Isotropic singularities in perfect fluid space-times were previously studied by \cite{GW,AT1} using the congruence of flow lines. In particular, \cite{GW} analysed the limiting behaviour of the physical space-time near the singularity when the latter is isotropic. It will be shown in Section \ref{6} how these limits can be directly obtained from boundedness conditions on the tractor curvature.

We use the $1+3$ decomposition with respect to a time-like congruence, which we outline in the next section. Our setup will follow that of \cite{EvE,UvE,WE}, so that for this paper (unlike \cite{article1}) we will adopt the signature convention $(-\, +\, +\, +) $, as used in this literature.

We make use of conformal densities and make some use of the tractor formalism, though most formulae are tensorial. For an introduction to tractors and the notation the reader is refered to \cite{article1}. In abstract indices, the conformal, physical and unphysical metric are written $\cg_{ab}, \gtilde_{ab}, g_{ab}$ respectively, so that $\gtilde_{ab}=\Omega^2g_{ab}$. To simplify notation we write all equations in the untilded, unphysical variables $( \D, g)$. However when particular physical assumptions simplify the equations, we will highlight this by using the physical variables $( \Dtilde, \gtilde) $.

\section{1+3 decomposition}
We review those parts of the formalism of $1+3$-decomposition with respect to a time-like congruence that we shall need, indicating the conformal properties of the familiar kinematic quantities associated with the congruence. Our main interest is in the spatially-homogeneous perfect-fluid cosmologies, when we use the fluid congruence, which is geodesic and twist-free in the physical metric. Since we shall only rescale with functions of $t$, it has these properties in the unphysical metric too, and it also turns out to be a congruence of conformal geodesics.

Given the time-like congruence with velocity field $v^a$ we define the \emph{velocity scale}  $v:=\mod\cg_{ab}v^av^b\mod^\half \sect(\e[1])$ in the notation of \cite{article1}; recall this means that $v$ is a conformal weight one scalar, a section of the line bundle $\Gamma(\e[1])$ of these. Then there is an associated unit velocity $u^a:=\frac{v^a}{v} \sect(\e^a[-1])$ and acceleration $A^b:= u^a \D_a u^b \sect(\e^b[-2])$, respectively conformal weight -1 and -2 vector fields. Given the application we have in mind, we will set the twist of the congruence to zero from the outset.

We have projection operators: $U\tensor{a}{b}{}=-u_a u^b$ onto the flow lines and $h\tensor{a}{b}{}=\d\tensor{a}{b}{}-U\tensor{a}{b}{}$ orthogonal to them; thus $U^2=U$, $h^2=h$ and $U \after h = h \after U =0$. The operator $h^{ab}_{cd}:=h\tensor{}{(a}{c} h\tensor{}{b)}{d} - \third h^{ab} h_{cd}$ projects out the spatial, symmetric, trace-free part of a tensor. All three operators have conformal weight zero and are used to decompose tensors into components along and orthogonal to the flow.

\noindent  With the twist-free assumption,  the decomposition of the covariant derivative $\D_a u_b$ takes the simplified form
\be
\label{Dudecomp}
  \D_a u_b = -u_aA_b+ \third \theta h_{ab} + \sigma_{ab}  \sect(\e_{ab}[1]) ,
\ee
in terms of the kinematic quantities:  the acceleration $A^b \sect(\e^b[-2])$, expansion $\theta \sect(\e[-1]) $, and shear $\sigma_{ab} \sect(\e_{(ab)}[1])$.

We define the magnitude $\sigma^2=\half \sigma_{ab}\sigma^{ab}$. The flow is hypersurface-orthogonal, with induced metric $h_{ab}$ and second fundamental form $\chi_{ab}=\third \theta h_{ab} + \sigma_{ab}$  on these hypersurfaces.

We recall (\cite{article1}) that a conformal density is \emph{homogeneous} if the change under a gauge transformation to a general Weyl connection $\Dhat = \D + b$ has no additive correction in the one-form $b_c$ which determines the change in connection. We can express the inhomogeneous conformal density $\theta$ in terms of an average length scale $L \sect(\e[1]) $  by
\be
\label{average length}
  \th = 3\frac{\D_u L }{L} \implies   \hat{\theta}=\theta +3(b,u) \sect(\e[-1]), \\
\ee
and then $L$ is a homogeneous density.


The Einstein tensor $G_{ab}$ is related to the Schouten tensor $P_{ab}$ by $${G}_{ab} := 2(P_{ab} - P_{cd}g^{cd} g_{ab}) $$ and in the physical space-time is equated to (a multiple of) the energy momentum tensor $\tilde{T}_{ab}$.
Under connection translation $\Dhat=\D+b $ the Schouten tensor transforms as
\be
\label{Schouten}
  P_{ab} - \Phat_{ab} = \D_a b_b - b_a b_b + \half \cg_{ab}\cg^{cd} b_c b_d.
\ee
In this article we will only use Levi-Civita connections, so that the 1-forms $b_a$ are always closed and the Schouten and Einstein tensors are always symmetric. In general, therefore, we define \emph{the geometric matter variables} by decomposing the Einstein tensor $G_{ab} $ according to
\be
\label{Tdecomp}
  G_{ab} = \mu u_a u_b +q_a u_b + u_a q_b + p h_{ab} + \pi_{ab},
\ee
where $\mu$ and $p$ are scalars, $q_a$ is a vector orthogonal to $u^a$ and $\pi_{ab}$ is a trace-free, symmetric tensor orthogonal to $u^a$. (In \cite{GW} the geometric matter variables were denoted by $A, B, \Sigma_a, \Sigma_{ab} $.)
In the physical space-time, the corresponding matter variables are tilded: $\mutilde$ is the energy density relative to $\tilde{u}^a$, $\ptilde$ is the isotropic pressure, and $\qtilde^a$  and $\pitilde_{ab}$ are the momentum density and the trace-free anisotropic pressure, both zero for a perfect fluid.
The trace-free Ricci tensor $^{(3)}S_{ab}  $  of the hypersurfaces can be written as
\be
  ^{(3)}S_{ab} = E_{ab} + \half \pi_{ab} - \third \th \sigma_{ab} + \sigma\tensor{a}{c}{}\sigma_{cb} - \frac{2}{3}\sigma^2 h_{ab} .
\label{3curvature}
\ee
where  $E_{ab}$ is the electric part of the Weyl tensor.

Finally in this section, for a perfect fluid the conservation equation implies
\be
\label{pf2ndBianchi}
\frac{\Dtilde_{\utilde} \mutilde}{(\mutilde+ \ptilde)} = - 3 \frac{\Dtilde_{\utilde} \Ltilde}{\Ltilde}
\ee
which can be solved for a polytropic perfect fluid, when $p=(\gamma-1)\rho$, to give
\be
\mutilde=\mutilde_* \Ltilde^{-3\c},\label{moo}
\ee
 where $\mutilde_*$ is constant along the flow.

\section{Bianchi space-times}

A space-time is said to be spatially homogeneous if it admits a 3-dimensional vector space of Killing vector fields (KVFs) transitive on space-like 3-surfaces. The KVFs form a Lie algebra with basis $\xi_\ap$ say, with $ {\alpha}'=1,2,3$, and the associated Lie group $G_3$ is the corresponding group of isometries.
The Bianchi models are classified in terms of the structure constants $C\tensor{}{\mup}{\ap\bp} $ of the Lie algebra, where $[\xi_\ap, \xi_\bp] = C\tensor{}{\mup}{\ap\bp}\xi_\mup $ (see e.g. \cite{WE,MC1})
and one then finds nine canonical forms, the Bianchi types.

We shall restrict to non-tilted perfect fluid cosmologies, for which the velocity $\utilde$ of the fluid (a unit vector for the physical metric) is orthogonal to these 3-surfaces, which can be labelled by proper-time $t$ along the fluid flow. Comoving coordinates $x^{\underline{\alpha}'}$, with $\underline{\alpha}'=1,2,3,$ (coordinate indices underlined) can be introduced in a standard way for each Bianchi type. We denote the Jacobi fields obtained from these standard comoving coordinates by $\eta_{\underline{\alpha}'}=\partial/\partial x^{\underline{\alpha}'}$ . From spatial homogeneity, we have $\Lie_\xi \utilde =0 $, that is the velocity is invariant under the action of the KVFs.
To construct a group-invariant orthonormal frame at one time, choose an orthonormal frame $\{\tilde{e}_\a^i \}$, $\alpha=0,1,2,3$, at one point on an orbit with $\tilde{e}_0^i=\utilde^i $ and Lie drag it around the 3-surface, so that $\Lie_\xi \tilde{e}_\a ^i= 0$. Then $\Lie_\xi \tilde{g}(\tilde{e}_\a, \tilde{e}_\b)=0$ so that the frame is orthonormal everywhere on the orbit.

For the propagation along the congruence, note that $\Lie_{[\tilde{u},\xi]}\tilde{e}^i_\a =0$ so that necessarily
\be\label{rot}
\Lie_{\tilde{u}} \tilde{e}^i_\a = M\tensor{\a}{\b}{}(t) \tilde{e}^i_\b ,
\ee
for some $M\tensor{\a}{\b}{}(t)$, which at this stage is freely specifiable.


In the next section we shall see that the fluid flow lines are conformal geodesics, when it is natural to choose a group-invariant frame which is Weyl-propagated, that is
\[\Dhat_u e_\a^i=0,\]
where $\Dhat$ is the Weyl connection defined from the conformal geodesic equation and $u^a$ is tangent to the conformally-parametrised conformal geodesics.  We shall take $u=e_0$.
  Then \[\Lie_u e^j_\a = \Dhat_{e_\a} u^j = e_\a^i  \th\tensor{i}{j}{},\] where $\th_{ij} $ is the unphysical second fundamental form which, by the spatial homogeneity, depends only on time.
The connection coefficients $\Gamma\tensor{}{\mu}{\a\b}$ of the connection  $\D$ in the chosen frame are defined by
\be
\label{connection coefficients}
  \D_{e_\b}e_\a^i =  \Gamma\tensor{}{\mu}{\a\b}e_\mu^i .
\ee

\section{The fluid flow-lines as conformal geodesics}

In the physical space-time the conformal geodesic equations can be written in a third-order form (given in \cite{article1}, but remembering that the signature here is opposite to the one used there): if the unit velocity vector is $\utilde^i$ with acceleration $\Atilde^i$ then
\be
 \label{ucg1}
  \tD\Atilde^i + (\utilde^k\tD\Atilde_k)\utilde^i = -\Ptilde\tensor{j}{i}{}\utilde^j - (\Ptilde_{jk}\utilde^j\utilde^k)\utilde^i .
\ee
where $\tD=\utilde^i\tilde\nabla_i$. The conformally-parametrised tangent vector is $u^i=q^2\utilde^i$ where $q$ is found from
\be
  \label{ucgq}
  2\frac{\tD^2 q}{q} = -\Ptilde_{jk}\utilde^j\utilde^k + \half \Atilde_j\Atilde^j ,
\ee
and the one-form $\tilde{b}_i$, associated with the congruence and used in Weyl propagation, is obtained as
\be
  \label{ucgb}
  \tilde{b}_i = -\Atilde_i + 2\frac{\tD q}{q} \utilde_i .
\ee
We set $\l=-2\frac{\tD q}{q}=\tilde{b}_j \utilde^j$.

In a (non-tilted) perfect fluid Bianchi space-time the matter flow lines are space-time geodesics and the Schouten tensor has the form
\be
\label{P in flow quantities}
  \Ptilde_{ij} = \left( \third \mutilde + \half \ptilde \right) \utilde_i \utilde_j + \frac{1}{6} \mutilde \tilde{h}_{ij}.
\ee
It is then straightforward to check that (\ref{ucg1})-(\ref{ucgb}) can be solved with $\Atilde_i=0$ and  $\btilde_i=-\l \utilde_i$. Thus the flow lines of a perfect fluid in a Bianchi space-time can be reparametrised as conformal geodesics. The 1-form $\btilde_i$ is exact and hence $\Dhat = \Dtilde + \btilde$, the Weyl connection associated with the congruence, is actually the Levi-Civita connection preserving the metric $g_{ab} =\Om^{-2}\gtilde_{ab}$ for $\Om=q^{2}$. In particular therefore the Weyl-propagated tetrad is parallelly-propagated in the unphysical metric. The conformal parameter $\tau$ along the conformal geodesics is given by
\[
  \t = \int^t \frac{1}{q^2} dt ,
\]
up to additive constant, where $t$ is proper-time in the physical metric. In the rescaled space-time, the conformally-parametrised tangent vector $u^i= q^2 \utilde^i $ is a unit vector, so that it is in fact affinely parametrised, and $\tau$ is the corresponding proper time in the unphysical metric.

For a perfect fluid with a polytropic equation of state and polytropic index $\c$, (\ref{ucgq}) takes the form
\be
\label{ddq}
  \ddot{q} = -\half \left(\third \mutilde + \half \ptilde\right) q = -\frac{3\c-1}{12}\mutilde q ,
\ee
with the overdot for $d/dt$. The freedom in the choice of solution to (\ref{ddq}) corresponds to the freedom in fractional-linear transformation of the conformal parameter.

Suppose that, into the past along each worldline, we reach the singularity as $t \to 0 $. The volume goes to zero in the physical metric, so for the volume to be nonzero in the rescaled space-time we require $q \to 0$ in that limit.

We next assume that there is a choice of $q$ such that $\tau$ does not have infinitely many poles before the singularity is reached, and is not automatically infinite at the singularity.  As discussed in \cite{article1}, this is an assumption on the singularity in the physical space-time which is logically prior to  that of bounded tractor curvature. Thus from now on we assume that $q$ vanishes at $t=0$ but not in some initial interval $(0, t_1)$, and that the integral for $\tau$
\be\label{taucondition}
  \t = \int_0^t \frac{1}{q^2} dt ,
\ee
is finite for $t\leq t_1$.

From the definition of $\l$, we find the expression $q= q_0 e^{-\half \int \l dt}$, so that $\l \to -\infty$ as $t \to 0$. Furthermore $\l q^2 = -2 q \tilde{D}q $ diverges near the singularity. To see this, suppose conversely that, on an initial interval $(0, \e)$, $ \tilde{D} q^2=-\l q^2$  is bounded, by $C$ say, then
\be
  q^2(t) \le  Ct\mbox{ so that } \int q^{-2}(t) dt  \ge \int \frac{dt}{Ct} \mbox{ on } (0, \e),
\ee
but this would violate our assumption on $\t$ after (\ref{taucondition}). These limits are needed in Section \ref{6}.

\section{The Global Extension Theorem and limits at the singularity}\label{6}

In \cite{article1} the following local extension theorem, for singularities to the future, was proven:
\begin{theorem}
\label{Maintheorem}
Let $\c:[0,\t_F) \to M$ be the final segment of an incomplete conformal geodesic in $(M,g)$, such that $b$ is bounded in $[0,\t_F) $. Let $W \subset M$ be a neighbourhood of $\c[0,\t_F) $ in which the strong causality condition holds. Let $\{e_\b \}$ be a Weyl propagated orthonormal frame along $\c$ with associated tractor frame $\TE_\TB $.

\noindent i) Suppose the tractor curvature $\TCurv$ and its derivative $\TD_v\TCurv $ along $\c$ have bounded norms on $\c$ with respect to $\{e_\b \}$ and $\TE_\TB $. Then there  exists a neighbourhood $U$ of $\c[0,\t_F)$ with $\overline{U} \subset W $ and a diffeomorphism $ V \subset \real^4 \to U $.

\noindent ii) Suppose that in $U$ the tractor curvature norms $\norm\TCurv^{(k+1)}\norm $  and $\norm\TCurv^{(1,k+1)} \norm$ are bounded. Then there exists a general Weyl connection $\Dhat$ associated to $[g]$ and a conformally related metric $g_{ij}$ such that there exists $U^*\supset \overline{U}$ with a $C^k $-extension of $(U, g_{ij}\rest_U) $ into $(U^*, g_{ij}\rest_{U^*}) $.

\noindent iii) The Riemann curvature of $g_{ij}$ is $C^{k-1} $.

\noindent Thus the conformal structure $(M,\cg) $ is locally extendible.
\end{theorem}
\noindent Definitions of tractor curvature and the norms used here are given in \cite{article1}, but the terms can be understood as follows:  the components of the tractor curvature $\TCurv$ are the Weyl tensor and the Cotton-York tensor; boundedness in the tractor frame is equivalent to boundedness in the Weyl-propagated tetrad; the norms are Riemannian sums-of-squares in the specified basis; and $\norm\TCurv^{(k+1)}\norm $  and $\norm\TCurv^{(1,k+1)} \norm$  are norms of $(k+1)$-fold derivatives of $\TCurv$ and their derivatives (once) along $u$.

In this article, we are considering initial rather than final singularities and hence we use initial segments $(0,\t_1]$. In \cite{article1} it was remarked that the extension theorem was only local because one could not guarantee that the set $\cal{O}$, the neighbourhood of $\gamma$ used for the extension process, included more than one singular point.  For Bianchi space-times, the spatial homogeneity ensures that all the conformal geodesics in the perfect fluid congruence are equivalent. Hence conditions satisfied by the chosen central curve $\c$ are satisfied by all its neighbours. In particular, this means the work underlying $(i)$ of Theorem \ref{Maintheorem} is avoided since the usual coordinatisation of Bianchi space-times provides a good coordinate system. The local extension theorem above will therefore give a global extension theorem when applied to perfect fluid Bianchi space-times.

The version of the Whitney Extension Theorem used in \cite{article1} was for sets in $\mathbb{R}^n$ which satisfied Whitney's Property $\mathcal{P}$. We needed bounds on the derivatives of order $k+1$ of the function $f$, which it is desired to extend, in order to obtain appropriately uniform bounds on derivatives of $f$ of order $k$; then the theorem provides a $C^k$-extension $F$ of $f$. In the spatially-homogeneous context, we only ever consider functions of the one variable $t$ and so we have a simpler, in fact elementary, extension theorem:

\emph{If $f\in C^k((0,\tau_1])$ with all derivatives bounded and having limits at zero, then $f$ admits an extension $F$ to $C^k((-\infty,\tau_1])$ which is in fact analytic on $(-\infty,0)$.}

This is obtained from the polynomial
\[F(t)=\Sigma^k_{j=0}\frac{t^j}{j!}f^{(j)}(0),\]
in $t<0$. Thus we may assume a lower degree of differentiability here.

Finally the assumption of strong causality was made in \cite{article1} in order to prove Property $\mathcal{P}$, so again we don't need to make it in this context.
\medskip

A global extension theorem with minimal conditions on the tractor curvature can be stated as follows:
\begin{theorem}
\label{tractorBianchi}
Suppose that we are given a perfect fluid Bianchi cosmology with an initial singularity, and a choice of $q$ and an initial interval for which the conformal parameter $\t$ is finite at the initial singularity. Suppose further that along one, and hence any, fluid flow-line the tractor curvature $\TCurv $ and its derivative $\TD_u \TCurv$ along $u$ are bounded in a $\btilde$-propagated frame. Then there exists a conformal rescaling such that the rescaled metric $g_{ij}$ and its inverse are bounded at the initial singularity and hence extendible. Furthermore the unphysical quantities
\be
  R\tensor{ab}{c}{d}, \,\th, \, \sigma_{ab}, \, \mu, \, p, \, \pi_{ab}, \, \Gamma^{\mu}_{\a\b}, \,^{(3)}R_{ab}
\ee
defining the curvature and connection, are all bounded, with limits at the singularity.
\end{theorem}
\Proofstart
The 1-form $\btilde$ is exact and hence the general Weyl gauge and the unit velocity gauge (unphysical gauge) coincide: $\Dtilde + \btilde = \Dhat = \D$.
The $\btilde$-propagated frame is parallelly propagated with respect to $\D$, the flow lines are metric geodesics of $g_{ab} $ and the vectors $\eta_{\underline{\alpha}'}$ are Jacobi fields.
By Theorem 4.1 and Lemma 4.7 of \cite{article1}, boundedness of $\TCurv $  implies boundedness of the metric in the coordinate basis and the kinematic quantities in the Weyl-propagated basis, and that these quantities have finite limits at the singularity. Theorem 3.4 of \cite{article1} now gives boundedness of the inverse metric in the coordinate basis, and consequently of the matrix relating the Weyl-propagated basis to the Jacobi fields, and its inverse.  These quantities also have limits at the singularity.

The Weyl tensor is part of the tractor curvature and so is bounded by assumption.  For the Schouten tensor (equivalently, the Ricci tensor) we recall the definition
$$Y_{abc}=2\nabla_{[a}P_{b]c},$$
of the Cotton-York tensor, which is bounded by the assumption on $\TCurv$. With results in hand, it is straightforward to integrate this to find that $P_{ab}$ is bounded, with a limit at the singularity.
Thus all of the unphysical Riemann tensor is bounded at the singularity. Now Lemma 4.11 of \cite{article1} with $j=1$ implies boundedness of $\nabla_\eta e_\b^i$ for all choices of $\eta$, and therefore also of $\nabla_{e_\a}e_\b^i$ (in fact one obtains boundedness of the tractor derivatives of the tractor frame). This gives boundedness of the connection coefficients $\Gamma^{\mu}_{\a\b}$ with a limit at the singularity.

Finally, if $\TD_u \TCurv$ is also bounded then the Weyl curvature, which at this stage we know to be bounded, also has a limit at the singularity. From this follows that the 3-curvature of the hypersurfaces, $^{(3)}R_{ab} $, is also bounded and has a limit at the initial singularity.
\Proofend

The unphysical space-time is well behaved and bounded as long as the tractor curvature is. We use this fact and combine it with the rescaling transformations for curvature and the geometric matter variables to get some information on the unphysical space-time.
\begin{prop}
\label{physicallimits}
Suppose the conditions of Theorem \ref{tractorBianchi} hold. Let $\varphi = \half q^4 \mutilde$ and $ \psi=\l q^2$ then the following limits hold as $t \to 0$:
\bea
\label{psivarphilimit}
\quad  \frac{\psi^2}{\varphi} \to \frac{2}{3} , \\
\label{densitylimit}  \tilde{\th} \to +\infty ,\, \quad \quad \frac{\mutilde}{\tilde{\th}^2} \to \third ,\, \quad \quad   \frac{\tilde{\sigma}^2}{\tilde{\th}^2} \to 0  , \\
\label{entropylimit} \frac{\tilde{C}_{abcd}\tilde{C}^{abcd}}{\tilde{R}_{ab}\tilde{R}^{ab}} \to 0 \\
\label{Weylcurvlimit}2 E_{\a\b}  - 3 \c\psi \sigma_{\a\b}  \to 0,\quad  H_{\a\b} \to 0 . 
\eea
\end{prop}

{\bf Remark}: The limits (\ref{densitylimit} -  \ref{entropylimit}) coincide with those derived for isotropic singularites in Theorem 3.2 and 3.3 of \cite{GW}.

\Proofstart
We use the boundedness of the curvature components to prove the limits above.
The conformal geodesic equation (\ref{ucg1}) gives us $P_{ab}v^a = 0$ and hence by (\ref{P in flow quantities}) $\third \mu + \half p =0$.
Combining $\btilde_a = -\l \tilde{u}_a = - \psi u_a $ with (\ref{Schouten})  we have
\bea
  \Ptilde_{ab} &=& P_{ab} - \Dtilde_a  \psi u_b - \psi \Dtilde_a u_b - \psi^2  u_a u_b - \half \psi^2 g_{ab} , \non \\
  \Dtilde_a u_b &= & \D_a u_b + \btilde_a u_b + u_a \btilde_b - g_{ab}\btilde_c u^c =  \th_{ab} - 2 \psi u_a u_b - \psi g_{ab} \non . 
\eea
Contracting the first equation with $h^{ab}= q^4 \tilde{h}^{ab}$, respectively $ h^{ab}_{cd} $, and substituting (\ref{P in flow quantities}) with the perfect fluid condition on the left hand side, we obtain
\bea
  \varphi &=&P \tensor{c}{c}{} - \psi \th + \frac{3}{2}\psi^2 , \label{phii}\\
\label{varphipsirelation}
  \half \pi_{ab} &=& h_{ab}^{cd}{} P_{cd} = \psi \sigma_{ab} ,
\eea
where (\ref{phii}) uses (\ref{ucgq}). For the conformal time parameter to be finite at the boundary we have shown that $\psi $ is required to diverge. Since $P_{ab} $ is bounded, it follows that $\pi_{ab} $ is bounded and, from  (\ref{varphipsirelation}) that  the unphysical shear vanishes as $t \to 0$.

From (\ref{phii}), we deduce that $\varphi $ must diverge and then obtain the limit (\ref{psivarphilimit}). Next we note the transformation of the following kinematic quantities under rescaling:
\bea
\theta &=&\Omega(\tilde{\theta}-3\tilde{u}^a\Upsilon_a)\label{the1}\\
\sigma_{ab}&=&\Omega^{-1}\tilde{\sigma}_{ab}\nonumber\\
\sigma^2&=&\Omega^2\tilde{\sigma}^2\label{sig2}
\eea
where now $\Omega=q^2$ and $\tilde{u}^a\Upsilon_a=-\lambda$ . Using the boundedness of $\th$ and (\ref{the1}) and substituting for $\l$ we get the first two of  (\ref{densitylimit}) as $t \to 0 $.  For the third we use (\ref{sig2}) in the form $\sigma^2=q^4\tilde{\sigma}^2$ and the observation, made after (\ref{taucondition}), that $\lambda q^2$ diverges.

For the quotient of curvatures (\ref{entropylimit}), we observe that $\tilde{C}_{abcd}\tilde{C}^{abcd}=q^{-8}C_{abcd}C^{abcd}$ and $C_{abcd}C^{abcd}$ is bounded, while $q^8\tilde{R}_{ab}\tilde{R}^{ab}=q^8(\mutilde^2 + 3\tilde{p}^2)\geq 4\varphi^2 $, and so diverges. Hence this ratio tends to zero.

For a polytropic perfect fluid in the physical space-time, (\ref{moo}) gives $\tilde{\mu} = \tilde{\mu}_* \tilde{L}^{-3\c} = \tilde{\mu}_* q^{-6\c} L^{-3\c}$. Since $\th$ is bounded, $L$  tends to a finite non-zero limit and therefore so does $q^{6\c} \tilde{\mu} $. The range $\c \le \frac{2}{3} $ and equation (\ref{varphipsirelation}) would now imply that $\varphi, \psi$ are bounded. As discussed earlier, this would prevent us from reaching the singularity in finite $\t$, so that this range in $\gamma$ has been ruled out by our assumptions (of course on physical grounds we are only interested in $1\le \c \le 2 $).

If $\c = \frac{4}{3}$ then $\tilde{\mu}q^8 $ has a finite non-zero limit. Then by (\ref{psivarphilimit}) $\l q^4 $ is finite too. Rewriting (\ref{varphipsirelation}) with the physical shear, $\l q^4 \tilde{\sigma}_{ab} =   h_{ab}^{cd}{} P_{cd} $ , we get a finite limit for $\tilde{\sigma}_{ab}$. Similarly if  $\c \ge \frac{4}{3}$ the physical shear must vanish at the singularity.

To analyse the behaviour of the Weyl curvature we look at the tractor curvature.
We assumed that $\TD_v \TCurv, \, \TCurv $ are bounded in the $\btilde$-propagated conformally orthonormal frame to be able to deduce the existence of a space-time extension through the initial singularity. It followed that $P_{ij}, C_{ijkl}$ and the Cotton-York tensor $ \CY_{ijk} $ are bounded. We have also shown that $\l$ and $\psi=\l q^2$ need to be unbounded.

In the physical space-time with perfect fluid we use (\ref{pf2ndBianchi}) to deduce that the physical Cotton-York tensor is given by
\be
  \widetilde{\CY}_{ijk}=2 \Dtilde_{[i}\Ptilde_{j]k} = -(\mutilde+\ptilde) \utilde_{[i}\stilde_{j]k} .
\ee
From the transformation rule of the Cotton-York tensor we obtain that $ \widetilde{\CY}_{ijl} + C\tensor{ij}{k}{l} \btilde_k=\CY_{ijl} $ is bounded. In the Weyl-propagated frame $e^i_\a$ we find
\bea
  \CY_{ijl}e_\a^i u^j e_\c^l &=& q^6 \left( \l \tilde{E}_{\a\c} - \half (\mutilde + \tilde{p}) \tilde{\sigma}_{\a\c} \right) \non \\
\label{CY1a}
  &=& \psi E_{\a\c} - \half q^4 (\mutilde + \tilde{p}) \sigma_{\a\c} , \\
\label{CY2a}
  \frac12\epsilon\tensor{pq}{ij}{}\CY_{ijl}e_\a^p u^q e_\c^l &=&  q^6\l \tilde{H}_{\a\c} = \psi H_{\a\c} .
\eea
Equation (\ref{CY1a}) relates the unphysical electric Weyl and shear tensor to the physical density and pressure. The boundedness of the left-hand-side of (\ref{CY1a}) with (\ref{psivarphilimit}) gives the first of (\ref{Weylcurvlimit}); the boundedness of the left-hand-side of (\ref{CY2a}) and the fact that $\psi$ diverges imply the second.

From (\ref{3curvature}) at $t=0$ and using these limits we obtain
\[^{(3)}S_{ab} = E_{ab} + \half \pi_{ab}=\frac{(3\gamma+2)}{3\gamma} E_{ab},\]
so that the trace-free 3-curvature of the hypersurfaces of homogeneity has a limit at the initial singularity, where it is determined by the initial electric Weyl tensor and vice versa.
\Proofend

These limits for isotropic singularities were found in \cite{GW} by assuming the existence of an initial isotropic singularity. We have derived them directly from the assumption of bounded tractor curvature and its derivative, and finiteness of the conformal time parameter, in the particular case of spatial homogeneity.

\section{Conclusion}

We have shown that perfect fluid flow lines in Bianchi space-times are conformal geodesics, which has allowed us to apply the Extension Theorem \ref{Maintheorem}. Then spatial homogeneity has resulted in a Global Extension Theorem \ref{tractorBianchi}, where we can extend across the whole singularity at once.
This theorem depends strongly on the spatial homogeneity of the Bianchi space-times and the fact that the perfect fluid flow lines can be written as conformal geodesics. It remains to be seen how one could derive such theorems for other matter models or for inhomogeneous space-times.

\end{document}